\documentstyle[aps,floats,epsfig,twocolumn]{revtex}

\begin{document}

\title{Non-Equilibrium Quasiclassical Theory for Josephson
Structures}
\author{M.~H.~S. Amin}
\address{D-Wave Systems Inc., 320-1985 West Broadway,
Vancouver BC, V6J 4Y3, Canada} \maketitle

\begin{abstract}
We present a non-equilibrium quasiclassical formalism suitable for
studying linear response ac properties of Josephson junctions. The
non-equilibrium self-consistency equations are satisfied, to very
good accuracy, already in zeroth iteration. We use the formalism
to study ac Josephson effect in a ballistic superconducting point
contact. The real and imaginary parts of the ac linear
conductance are calculated both analytically (at low frequencies)
and numerically (at arbitrary frequency). They show strong
temperature, frequency, and phase dependence. Many anomalous
properties appear near $\phi = \pi$. We ascribe them to the
presence of zero energy bound states.
\end{abstract}

\section{Introduction}

Quasiclassical theories are proven to be powerful for studying
superconducting systems. They have been used by many researchers
to study the properties of superconductors in the Meissner
\cite{Mstate} or vortex states \cite{Maki,IHEM,RSW}, as well as
for surfaces \cite{MS,Rainer}, point contacts
\cite{KO,Zaitsev,KO2,LY}, or grain boundaries between different
superconductors \cite{Barash,Yip,amin,2D}. The equilibrium
quasiclassical theory was developed by Eilenberger
\cite{Eilenberger} and Larkin and Ovchinnikov \cite{LO1} by
integrating out irrelevant small scale degrees of freedom from
the Nambu-Gorkov Green's function formulation of BCS
superonductivity \cite{Gorkov}. It was later generalized to
non-equilibrium by Eliashberg \cite{Eliashberg} and Larkin and
Ovchinnikov \cite{LO2}.

A major development in the numerical calculations in equilibrium
quasiclassical theory was made after the introduction of
Riccati-transformation by Schopohl and Maki \cite{Maki,Schopohl}.
The transformation changes the Eilenberger equations
\cite{Eilenberger} into a set of decoupled non-linear
differential equations, that can be integrated easily. In
non-equilibrium, the presence of convolution integrals in the
equations of motion for the Green's functions makes the formalism
nontrivial. Nevertheless, a generalized version of the
Schopohl-Maki transformations for non-equilibrium systems has
been suggested by Eschrig {\em et al}. \cite{eschrig,eschrig2}.

Josephson junctions are important devices, not only due to their
rich physical properties, but also for many applications,
including sensitive magnetometers \cite{Kirtley,AOZ}, ultrafast
switching devices \cite{Likharev}, qubit prototypes
\cite{MSS,ftq}, etc. dc and ac properties of them have been the
subject of extensive research
\cite{Likharev2,GZ,Bratus,AB,AB2,ZA}. Some of the investigations
are based on the tunneling Hamiltonian approach \cite{tinkham},
which provides a good approximation when the transparency of the
junction is small (e.g.~tunnel junctions). At large
transparencies, which is the case for superconducting point
contacts or grain boundary junctions, multiple Andreev
reflections (MAR) take place \cite{KBT}. The MAR theories work
well when the biasing voltage is large. At small biasing
voltages, the number of Andreev reflections grows ($\sim
\Delta/eV$, with $\Delta$ being the superconducting order
parameter). Nevertheless, the formalism was applied to the case
of single channel superconducting quantum point contact with
small biasing voltage \cite{AB}. Alternatively, non-perturbative
Hamiltonian method \cite{LY} and non-equilibrium Green's function
method \cite{ZA} were developed. However, most of these theories
are suitable only when the applied voltage is constant. An
exception is Ref.~\cite{AB2}, which studies a superconducting
quantum point contact in the presence of an ac voltage, but only
in adiabatic regime. Thus a theory capable of studying high
transparency Josephson junctions with an ac biasing voltage at
arbitrary frequency is still lacking.

For equilibrium systems, the quasiclassical Green's functions
theory has provided a convenient tool to study Josephson
structures with arbitrary transparency and roughness of the
junctions (see Ref.~\cite{amin} and references therein). A
generalization of the formalism to the non-equilibrium case, may
also provide a powerful and convenient method to study ac
properties of such structures. In this article, we rewrite the
theory of Refs.~\cite{eschrig,eschrig2} in a form suitable for
studying properties of a {\em general} Josephson junction in the
presence of an ac voltage at {\em arbitrary} frequency. To our
knowledge, no such theory exists. We apply the theory to the case
of a ballistic point contact between two conventional ($s$-wave)
superconductors. At low frequencies, we find closed analytical
expressions for the real and imaginary parts of the ac
conductivity. They agree very well with the numerical results,
except where the low frequency expansion fails. Both quantities
show strong temperature, frequency, and phase dependence. We
observe anomalous behavior when the phase difference across the
point contact approaches $\pi$. We relate that to the presence of
the zero energy bound states.

Section II introduces the formalism and formulates it in a form
appropriate for studying Josephson systems. Section III is
devoted to calculation of ac current through a ballistic point
contact between two $s$-wave superconductors. The low frequency
analytical results are given in III-A. The numerical results, as
well as a comparison with the analytical results, are presented
in III-B. Section IV summarizes the main results. A detailed
description of the theory and notations is given in two
appendices. Understanding the appendices in great detail is not
necessary for understanding the main body of the article and for
application of the theory to other problems.

\section{The formalism}

A convenient way to study non-equilibrium systems is to use
Keldysh Green's functions \cite{Zbook}. The quasiclassical
approximation of the Keldysh formalism introduces Retarded,
Advanced, and Keldysh Green's functions: $\widehat{g}^\alpha$,
$\alpha = R,A,K$. The first two describe spectral distribution of
the states of the system, while the latter has information about
population of those states.  They are all 2$\times$2 matrices,
and functions of the Fermi velocity ${\bf v}_F$, quasiparticle
energy $\epsilon$, position ${\bf r}$, and time $t$. The exact
definitions of $\widehat{g}^\alpha$ and their equations of motion
are given in appendix A [see Eqs.~(\ref{defg}) and (\ref{QCEq})].
The method we present here is a linear response treatment of
$\widehat{g}^\alpha$. We consider a clean superconducting system
in the absence of an external magnetic field. The coupling to the
electromagnetic field is via the vector potential ${\bf A}$ and
scalar potential $\Phi$. As will become soon clear, working with
a gauge in which ${\bf A}=0$ simplifies the calculations
significantly. In such a gauge, $\Phi$ is the only perturbation
applied to the system, which we take to be small.

Let us first consider the case of a uniform (bulk)
superconductor. We introduce a gauge transformation (see appendix
A for details) $\Delta \mapsto \widetilde{\Delta}= e^{-i\delta
\phi} \Delta$. Under such a transformation $e\Phi \mapsto e\Phi -
(1/2)\partial_t \delta \phi $. (Throughout this article we use
$\hbar=k_B=1$.) We choose $\delta \phi$ in such a way to exactly
eliminate $\Phi$ from the gauge transformed dynamical equations
[Eq.~(\ref{QCEq})]. Thus, $\delta \phi$ should satisfy
\begin{equation}
{\partial \delta \phi \over \partial t} = 2e\Phi,  \label{JosR}
\end{equation}
which is the well-known Josephson relation \cite{ESA}. The vector
potential ${\bf A}$ is still zero after this gauge
transformation, because $\nabla \delta \phi = 0$ within the bulk.
The gauge transformed equation of motion is therefore exactly the
same as the equilibrium equation. As a result, $\widetilde{\Delta}
= \Delta_0$, or $\Delta = \Delta_0 e^{i\delta \phi}$, where
$\Delta_0$ is the equilibrium order parameter. This means that
the magnitude of the order parameter, $|\Delta|$, is independent
of $\Phi$, whereas its phase varies with $\Phi$ via the Josephson
relation (\ref{JosR}). In other words, if the only perturbation
to the system is through the time varying potential $\Phi$, its
only effect is to change the phase of the order parameter
\cite{dlt}.

The above simple observation has very important consequence in
our linear response formalism. One can write the original (gauge
dependent) order parameter as

\begin{equation}
\Delta = \Delta_0 e^{i \delta \phi} \approx \Delta_0 (1 + i
\delta \phi) = \Delta_0 + \delta \Delta,
\end{equation}
where $\delta \Delta$ is the non-equilibrium linear response
correction to the order parameter. In Fourier space, the
Josephson relation (\ref{JosR}) becomes $\delta \phi = i2e
\Phi(\omega)/\omega$, which yields

\begin{equation}
\delta \Delta (\omega) = -{2 \Delta_0 \over \omega} e\Phi
(\omega), \label{dd}
\end{equation}
This value indeed satisfies the self-consistency equation for the
homogeneous superconductor [cf. Eq.~(\ref{SCEq})]. To ensure
small perturbation requirement, one needs $e \Phi \ll \omega$.
Charge neutrality condition in the bulk is also satisfied
automatically by using (\ref{dd}). This can be seen simply from
the above gauge transformation argument, taking into account the
neutrality of the unperturbed (equilibrium) system.

We now consider a Josephson junction with an equilibrium phase
difference $\phi$ and an ac voltage $V=V_0 \cos \omega t$ across
the junction. Far away from the junction, we take the phase of the
order parameter and the scalar potential on the left (L) and
right (R) sides to be $\phi_{\rm R,L} = \pm \phi/2$ and
$\Phi_{\rm R,L} = \pm (V_0/2)\cos \omega t$, respectively. In
general, the phase of the order parameter is space dependent.
Therefore, performing the above gauge transformation will produce
a vector potential ${\bf A} = (c/2e) \nabla \delta \phi$
($\propto j_{\rm ac}/j_{\rm c, bulk}$, where $j_{\rm ac}$ is the
ac current density in the banks and $j_{\rm c, bulk}$ is the bulk
critical current density), which invalidates our arguments.
However, in most practical systems, $j_{\rm ac} \ll j_{\rm c} \ll
j_{\rm c, bulk}$, where $j_{\rm c}$ is the Josephson current
density. The corrections to Eq.~(\ref{dd}) are therefore small
and $O(j_{\rm c}/j_{\rm c, bulk})$ \cite{jac}. Thus one can still
use (\ref{dd}), as a very good approximation, even when
$\Delta_0$ is space dependent. This removes the necessity for an
iterative procedure (the main obstacle in these types of
calculations) in order to self-consistently calculate $\delta
\Delta$, and satisfy charge neutrality condition (within the
bulk) \cite{eschrig,eschrig2}. The equilibrium order parameter
$\Delta_0$, however, should be calculated self-consistently using
the common iterative methods; convergence of such calculations is
proven to be very good, especially when using the Matsubara
technique \cite{amin}.

\subsection{Equilibrium Solution}

In equilibrium, the retarded and advanced Green's functions can
be written in terms of the Riccati amplitudes $a_0^\alpha$ and
$b_0^\alpha$ in a way very similar to the conventional method for
the Matsubara Green's functions \cite{amin}.

\begin{equation}
g_0^\alpha= s^\alpha {1 - a_0^\alpha b_0^\alpha \over 1 +
a_0^\alpha b_0^\alpha} \ , \qquad f_0^\alpha= s^\alpha {2
a_0^\alpha \over 1 + a_0^\alpha b_0^\alpha } \ , \label{gf0RA}
\end{equation}
where

\begin{equation}
s^\alpha = \left\{ \begin{array}{cc} + & \ {\rm for} \ \alpha=R \\
- & \ {\rm for} \ \alpha=A \end{array} \right. . \label{s}
\end{equation}
The subscript ``0'' denotes equilibrium quantities. The Riccati
amplitudes satisfy the following Riccati equations \cite{note}

\begin{eqnarray}
{\bf v}_{F}\cdot \nabla a_0^\alpha &=& 2i \epsilon^\alpha
a_0^\alpha -
(a_0^\alpha)^{2} \Delta^*_0 +  \Delta_0 , \nonumber \\
-{\bf v}_{F}\cdot \nabla b_0^\alpha &=& 2i \epsilon^\alpha
b_0^\alpha - (b_0^\alpha)^{2}\Delta_0 + \Delta^*_0 , \label{rcti}
\end{eqnarray}
where $\epsilon^\alpha = \epsilon + i s^\alpha \eta$, with
$\epsilon$ and $\eta$ being the real and imaginary parts of the
quasiparticle energy respectively. $\eta$ is the quasiparticle
damping, related to the inelastic lifetime $\tau$ of the
quasiparticles by $\eta = 1/\tau$ \cite{eta0}. Notice that the
scalar potential $\Phi$ does not appear in the equilibrium
equations. The boundary conditions are the bulk solutions of
Eq.~(\ref{rcti}):

\begin{equation}
a_0^\alpha = {\Delta_0 \over -i\epsilon^\alpha + s^\alpha
\Omega^\alpha}\ , \qquad b_0^\alpha = {\Delta^*_0 \over
-i\epsilon^\alpha + s^\alpha \Omega^\alpha} \ , \label{eqbc}
\end{equation}
where $\Omega^\alpha {=} \sqrt{|\Delta_0|^2 -
(\epsilon^{\alpha})^2 }$. The differential equations (\ref{rcti})
and the boundary conditions (\ref{eqbc}) can be obtained from
their counterparts in Matsubara formalism (see Ref.~\cite{amin}
for example), by changing $\omega_n \rightarrow
-i\epsilon^\alpha$, where $\omega_n$ is the Matsubara frequency.

To calculate the Riccati amplitudes at other points, one should
define quasiclassical trajectories as straight lines in the
direction of ${\bf v}_F$ (see Fig.~\ref{fig1}). $a_0^R$ and
$b_0^A$ are obtained by integrating (\ref{rcti}) in the direction
of ${\bf v}_F$ along the trajectory, starting from the boundary
conditions (\ref{eqbc}) at $-\infty$. For $b_0^R$ and $a_0^A$,
integrations are taken in the opposite direction. The following
symmetries exist for the equilibrium functions

\begin{equation}
a_0^A = - \left( b_0^{R} \right)^* , \qquad b_0^A = - \left(
a_0^{R} \right)^*. \label{symab}
\end{equation}
It is therefore sufficient to calculate one of the sets of
retarded or advanced functions.

In equilibrium, the Keldysh Green's function is related to the
retarded and advanced ones by

\begin{equation}
\widehat{g}_0^K = (\widehat{g}_0^R - \widehat{g}_0^A) {\cal F},
\label{gKRA}
\end{equation}
where

\begin{equation}
{\cal F} \equiv \tanh \left( {\epsilon \over 2 T} \right)
\end{equation}
takes into account the thermal distribution of the quasiparticles.
Equilibrium current and charge densities are calculated by
Eqs.~(\ref{j}) and (\ref{rho}), summing over all trajectories.

\subsection{Linear Response Solution} \label{LR}

Generalization of the above Riccati-transformation to
non-equilibrium is discussed in appendix B. The presence of the
$\otimes$-operations [see Eq.~(\ref{otimes}) for definition]
makes calculations nontrivial. However, significant
simplification arises when one side of the $\otimes$-operation is
an equilibrium quantity (and therefore time/frequency
independent). More specifically, in frequency space we have

\begin{eqnarray}
P_0 (\epsilon) \otimes Q(\epsilon,\omega) &=& P_0 \left( \epsilon
+ {\omega \over 2} \right) Q(\epsilon,\omega) , \nonumber \\
P(\epsilon,\omega) \otimes Q_0 (\epsilon) &=& P(\epsilon,\omega)
Q_0 \left(\epsilon - {\omega \over 2} \right).
\end{eqnarray}
To take the advantage of this property, we use linear expansion.
In other words, we assume that the perturbation to the system
(i.e.~$\Phi$) is so small that the linear expansion of the Green's
functions provides a good approximation to the exact solution.
All the differential equations [Eqs.~(\ref{dab}) and
(\ref{dabK})] will then be linear in the time-varying parts, and
Fourier transformation will be straightforward: no complicated
convolution integrals arise.

Let us introduce simplifying notation

\begin{equation}
\epsilon_\pm = \epsilon \pm {\omega \over 2}, \qquad Q_{0\pm} =
Q_0 \left(\epsilon_\pm \right).
\end{equation}
We define linear response Green's functions as $\delta
\widehat{g}^\alpha = \widehat{g}^\alpha - \widehat{g}_0^\alpha$,
where $\widehat{g}_0^\alpha$ are the equilibrium Green's
functions. Similarly, we introduce small corrections to the
Riccati amplitudes $\delta a^\alpha = a^\alpha - a_0^\alpha$ and
$\delta b^\alpha = b^\alpha - b_0^\alpha$. The linear response
Green's functions are then given in terms of the Riccati
amplitudes by

\begin{eqnarray}
\delta g^\alpha &=& -2 s^\alpha {\delta a^\alpha b_{0-}^\alpha +
\delta b^\alpha a_{0+}^\alpha \over (1 + a_{0+}^\alpha
b_{0+}^\alpha)(1 + a_{0-}^\alpha b_{0-}^\alpha)} \ , \label{dgRA} \\
\delta f^\alpha &=& 2 s^\alpha {\delta a^\alpha  - \delta b^\alpha
a_{0+}^\alpha a_{0-}^\alpha \over (1 + a_{0+}^\alpha
b_{0+}^\alpha)(1 + a_{0-}^\alpha b_{0-}^\alpha)} \ , \label{dfRA}
\end{eqnarray}
for $\alpha = R,A$.

We also define an anomalous Green's function $\delta
\widehat{g}^X$ by

\begin{eqnarray}
\delta \widehat{g}^K = \delta \widehat{g}^X ({\cal F}_+ - {\cal
F}_-) + \delta \widehat{g}^R {\cal F}_- - \delta \widehat{g}^A
{\cal F}_+ . \label{gx}
\end{eqnarray}
Correspondingly, we introduce anomalous functions $\delta a^X$ and
$\delta b^X$ [see Eqs.~(\ref{mdgx}) and (\ref{abx})] which are
related to the Green's functions through
\begin{eqnarray}
\delta g^X &=& 2 {\delta a^X - \delta b^X a_{0+}^{R} b_{0-}^{A}
\over (1 + a_{0+}^{R} b_{0+}^{R})(1 + a_{0-}^{A} b_{0-}^{A})}\ , \label{dgX} \\
\delta f^X &=& 2 {\delta a^X a_{0-}^{A} + \delta b^X a_{0+}^{R}
\over (1 + a_{0+}^{R} b_{0+}^{R})(1 + a_{0-}^{A} b_{0-}^{A})}\ .
\label{dfX}
\end{eqnarray}
The differential equations describing $\delta a^\alpha$ and
$\delta b^\alpha$ ($\alpha = R,A,X$) have general forms

\begin{eqnarray}
{\bf v}_F \cdot \nabla \delta a^\alpha &=& A^\alpha \delta
a^\alpha + B^\alpha , \nonumber \\
-{\bf v}_F \cdot \nabla \delta b^\alpha &=& \widetilde{A}^\alpha
\delta b^\alpha + \widetilde{B}^\alpha , \label{DifEqs}
\end{eqnarray}
with $A$'s and $B$'s given by \cite{note}

\begin{eqnarray}
A^\alpha &=& 2i \epsilon - \Delta_0^* (a_{0+}^\alpha
+ a_{0-}^\alpha),  \nonumber \\
B^\alpha &=& \delta \Delta + a_{0+}^\alpha a_{0-}^\alpha \delta
\Delta^* -ie\Phi (a_{0+}^\alpha - a_{0-}^\alpha), \label{AB}
\end{eqnarray}

\begin{eqnarray}
\widetilde{A}^{\alpha} &=& 2 i\epsilon - \Delta_0 (b_{0+}^\alpha
+ b_{0-}^\alpha), \nonumber \\
\widetilde{B}^{\alpha} &=& - \delta \Delta^* - b_{0+}^\alpha
b_{0-}^\alpha \delta \Delta + ie\Phi (b_{0+}^\alpha -
b_{0-}^\alpha), \label{ABT}
\end{eqnarray}
for the retarded ($\alpha=R$) and advanced ($\alpha=A$) functions,
and

\begin{eqnarray}
A^X &=& i\omega - a_{0+}^R  \Delta_0^* + b_{0-}^A \Delta_0,  \nonumber \\
B^X &=& -a_{0+}^R \delta \Delta^* +  b_{0-}^A \delta \Delta -
ie\Phi (1 + a_{0+}^R  b_{0-}^A), \label{ABX}
\end{eqnarray}

\begin{eqnarray}
\widetilde{A}^{X} &=& i\omega -  b_{0+}^R \Delta_0 + a_{0-}^A
\Delta_0^*, \nonumber \\
\widetilde{B}^{X} &=& b_{0+}^R \delta \Delta - a_{0-}^A \delta
\Delta^* + ie\Phi (1 + b_{0+}^R a_{0-}^A),
\end{eqnarray}
for the anomalous ones ($\alpha = X$). $\delta \Delta \equiv
\Delta - \Delta_0$, is given by Eq.~(\ref{dd}) and satisfies the
self-consistency equation

\begin{eqnarray}
\delta \Delta ({\bf v}_F) = {N_F \over 4i}
\int_{-\epsilon_c}^{\epsilon_c} d\epsilon \left< V({\bf v}_F,{\bf
v}_F') \delta f^K ({\bf v}_F') \right>_{{\bf v}_F'}, \label{SCEq}
\end{eqnarray}
where $V({\bf v}_F,{\bf v}_F')$ is the interaction potential,
$N_F$ the density of states at the Fermi surface, and
$\epsilon_c$ the energy cutoff. The bulk boundary conditions for
the amplitudes are

\begin{equation}
\delta a^\alpha = -{B^\alpha \over A^\alpha} \ ,\qquad \delta
b^\alpha = -{\widetilde{B}^\alpha \over \widetilde{A}^\alpha}.
\label{bc}
\end{equation}
To assure stability of the integrations, Eqs.~(\ref{DifEqs})
should be integrated along the trajectory in the direction of
${\bf v}_F$ for $\delta a^R$, $\delta b^A$, and $\delta a^X$, but
in the opposite direction for $\delta b^R$, $\delta a^A$, and
$\delta b^X$ (see Fig.~\ref{fig1}). Having the Green's functions,
the non-equilibrium correction to the current density is given by

\begin{equation}
\delta {\bf j} = {e N_F \over 4} \int_{-\epsilon_c}^{\epsilon_c}
d\epsilon \left< {\bf v}_F\ {\rm Tr} [\widehat{\tau}_3 \delta
\widehat{g}^K ] \right>, \label{dj}
\end{equation}
and the charge density is found from

\begin{equation}
\delta \rho = eN_F \left( -2e\Phi + {1 \over 4}
\int_{-\epsilon_c}^{\epsilon_c} d\epsilon \left< {\rm Tr} [\delta
\widehat{g}^K ] \right> \right). \label{drho}
\end{equation}

\section{ac Josephson effect in a
superconducting point contact}

\begin{figure}[t]
\epsfysize 4.7cm \epsfbox[50 320 400 600]{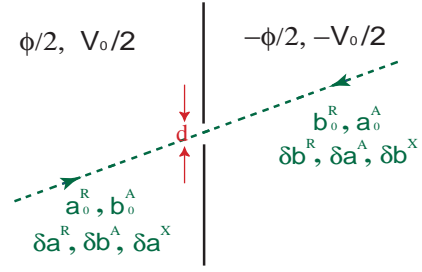} \caption{Two
superconducting regions connected via an orifice. The dashed line
shows the quasiclassical trajectory. The arrows indicate the
directions of integration.} \label{fig1}
\end{figure}

A ballistic superconducting point contact is probably the simplest
system that the method described here can be applied to. It
brings extra simplicity because even the equilibrium solutions
can be found non-self-consistently \cite{KO}. Nevertheless, the
system shows rich and nontrivial physical behavior. Let us
consider an orifice between two conventional ($s$-wave)
superconductors (Fig.~\ref{fig1}), dimension of which much
smaller than the inelastic scattering length and coherence length
of the superconductors ($d \ll l_\tau {=} v_F \tau, \xi_0 {=}
v_F/\pi |\Delta_0|$). We assume perfect transparency at the
contact, although generalization to arbitrary transparency is
straightforward \cite{amin,eschrig2}. We take the equilibrium
order parameter to be constant on both sides of the contact,
$\Delta_{\rm L,R} = |\Delta_0| e^{i \phi_{\rm L,R}}$, with
$\phi_{\rm L,R} = \pm \phi/2$, where $\phi$ is the phase
difference between the two sides. Here, L (R) denotes the left
(right) side of the contact.

We also take $\delta \Delta$ to be constant on each side of the
contact, given by Eq.~(\ref{dd}). The scalar potential $\Phi$ is
taken to be $\Phi_{\rm L,R} = \pm (V_0/2) \cos \omega t$ on the
left and right sides respectively, $V_0$ being the amplitude of
the potential difference. It is taken to be a real number and so
small that the linear expansion provides a good approximation
($eV_0 \ll |\Delta_0|,\omega$).

To find the current response, $I$, of the system, we calculate the
current density $\delta j_z$ at the orifice, using
Eq.~(\ref{dj}), and then integrate it over the area $S$ of the
orifice. In Fourier space, $I(\omega)$ can be a complex number.
The real part of it describes the dissipation of the system, while
its imaginary part gives information about inductive or
capacitive behavior of the system. The linear admittance of the
system is defined by $Y(\omega)= I(\omega)/V_0$ ($=1/Z$, where
$Z$ is the impedance of the contact).

We proceed with the calculation of the current in two different
ways. First, we find analytical results in the regime of small
$\omega$ and $\eta$. We then provide the results of full
numerical calculation and compare them with the analytical ones.

\subsection{Low Energy Analytical Results}

In the case of a point contact, it is not difficult to obtain
analytical results. Since the superconductor is homogeneous
everywhere except near the contact, the solution to the functions
$a$ and $b$ at the contact is almost equal to their bulk values.
More specifically, $a^R(0) = a^R(-\infty)$, $b^R(0) =
b^R(+\infty)$, etc. From now on we drop the arguments and just
write $a^R$, $b^R$, etc., keeping in mind that what we mean is
the values at the position of the contact. Substituting these
values in the corresponding equations, one can obtain analytical
expressions for the current. The exact expression is rather
complicated and does not give any more insight than the numerical
results. It, however, can be significantly simplified in low
energy regime.

Here, we calculate the contact admittance in the regime
$\eta,\omega \ll |\Delta_0|,T$ (but of course $\omega \gg eV_0$).
Let us first introduce the following parameterization

\begin{equation}
\epsilon^\alpha = |\Delta_0| \cos \gamma^\alpha, \qquad \alpha =
R,A
\end{equation}
where $\gamma^\alpha$ is a complex number. We therefore find
$\Omega^\alpha = |\Delta_0| \sin \gamma^\alpha$. This choice of
notation significantly simplifies the form of $a_0$ and $b_0$.
For instance from (\ref{eqbc}), taking the phase of the order
parameter to be $+\phi/2$ (left side of the contact), one finds

\begin{equation}
a_0^R ={|\Delta_0| e^{i\phi/2} \over -i\epsilon^R + \Omega^R} =
ie^{i(\phi/2 - \gamma^R)} = ie^{-i \delta \gamma^R},
\end{equation}
where $\delta \gamma^R \equiv \gamma^R - \phi/2$. Similarly,
$b_0^R$ has to be calculated on the right side of the contact
(where the phase is $-\phi/2$) and turns out to be exactly equal
to $a_0^R$. In general, one can write
\begin{equation}
b_0^\alpha=a_0^\alpha=ie^{-i s^\alpha \delta \gamma^\alpha}.
\end{equation}
First notice that $1+a_\pm^\alpha b_\pm^\alpha = 1 -
e^{-2is^\alpha \delta \gamma_\pm^\alpha}$ vanishes as $\delta
\gamma_\pm^\alpha \equiv \delta \gamma^\alpha
(\epsilon_\pm^\alpha) \rightarrow 0$ (or as $\gamma_\pm^\alpha
\rightarrow \phi/2$). Because of such expressions in the
denominators of Eqs.~(\ref{dgRA}) and (\ref{dgX}), the most
important contribution to those equations must come from points
close to $\epsilon = \epsilon_0(\phi)$ for which $\delta
\gamma^\alpha \ll 1$. Here
\begin{equation}
\epsilon_0 (\phi) = |\Delta_0| \cos {\phi \over 2}.
\end{equation}
($\pm \epsilon_0$ are indeed the energies of the Andreev bound
states in the contact \cite{Zbook}.) We can therefore focus only
on those points. Expanding the numerators of (\ref{dgRA}) and
(\ref{dgX}) up to first order in $\delta \gamma^\alpha$, (i.e.
first order in $\omega$ and $\eta$), we find
\begin{eqnarray}
\delta g^R &=& {eV \over \omega} \left[{1 \over \delta \gamma_+^R
\delta \gamma_-^R } +  \left( {\epsilon_0 \over 2\Omega_0} - i
\right) \left( {1 \over \delta \gamma_+^R} - {1 \over \delta
\gamma_-^R} \right) \right] \nonumber \\
\delta g^A &=& {eV \over \omega} \left[{1 \over \delta \gamma_+^A
\delta \gamma_-^A } +  \left( {\epsilon_0 \over 2\Omega_0} + i
\right) \left( {1 \over \delta \gamma_+^A} - {1 \over \delta
\gamma_-^A} \right) \right] \nonumber \\
\delta g^X &=& {eV \over \omega} \left[{1 \over \delta \gamma_+^R
\delta \gamma_-^A } +  \left( {\epsilon_0 \over 2\Omega_0} + i
\right) {1 \over \delta \gamma_+^R} \right. \nonumber \\
&& \left. \quad + \left( {\epsilon_0 \over 2\Omega_0} - i \right)
{1 \over \delta \gamma_-^A} \right], \label{dgRAX}
\end{eqnarray}
where
\begin{equation}
\Omega_0 (\phi) = |\Delta_0| \sin {\phi \over 2}.
\end{equation}
Let us write $\epsilon = \epsilon_0 + \epsilon'$ ($\epsilon' \ll
|\Delta_0|$). For $\phi \ne 0$ (we will discuss the $\phi=0$ case
later), one can write
\begin{equation}
\delta \gamma^\alpha_\pm \approx - {1 \over \Omega_0} \left(
\epsilon' + is^\alpha \eta \pm {\omega \over 2} \right).
\label{gammaexpnd}
\end{equation}
Therefore
\begin{equation}
{1 \over \delta \gamma^\alpha_\pm} \approx - \Omega_0 \left[ {P
\over \epsilon' \pm \omega/2} - i \pi s^\alpha \ \delta \left(
\epsilon' \pm {\omega \over 2} \right) \right],
\end{equation}
where $P$ gives the principal value integral when integrating
over $\epsilon'$. Because of symmetry, the principle value
integrals are negligible after integration. We therefore write
\begin{equation}
{1 \over \delta \gamma^\alpha_\pm} \approx i \pi s^\alpha
\Omega_0 \ \delta \left( \epsilon' \pm {\omega \over 2} \right).
\label{gm1}
\end{equation}
On the other hand
\begin{eqnarray}
{1 \over \delta \gamma^\alpha_+ \delta \gamma^\alpha_-} &=&
{\Omega_0^2 \over \omega} \left( {1 \over \epsilon' - \omega/2 +
i s^\alpha \eta} - {1 \over
\epsilon' + \omega/2 + i s^\alpha \eta} \right) \nonumber \\
&\approx& {i \pi s^\alpha \Omega_0^2 \over \omega} \left[ \delta
\left( \epsilon' + {\omega \over 2} \right) -  \delta \left(
\epsilon' - {\omega \over 2} \right) \right]. \label{gm2}
\end{eqnarray}
Similarly
\begin{eqnarray}
{1 \over \delta \gamma^R_+ \delta \gamma^A_-} &=& {\Omega_0^2
\over \omega + 2i\eta} \left( {1 \over \epsilon' - \omega/2 - i
\eta} - {1 \over
\epsilon' + \omega/2 + i \eta} \right) \nonumber \\
&\approx& {i \pi \Omega_0^2 \over \omega + 2 i\eta} \left[ \delta
\left( \epsilon' + {\omega \over 2} \right) +  \delta \left(
\epsilon' - {\omega \over 2} \right) \right]. \label{gm3}
\end{eqnarray}
Substituting (\ref{gm1})-(\ref{gm3}) into (\ref{dgRAX}) and using
(\ref{gx}), one can calculate $\delta g^K$ and thereby the total
current using (\ref{dj}). Taking the integral over $\epsilon'$ in
(\ref{dj}), expanding the resulting hyperbolic tangents around
$\epsilon_0$ (to the first order in $\omega/T \ll 1$) and keeping
only the leading order terms, we find the admittance to be
\begin{eqnarray}
Y(\omega) &=& {\pi \over \omega R_N} \left[ {\Omega_0^2  \over 2T}
\left( {2 \eta \over \omega + 2i\eta}\right)
{\rm sech}^2 {\epsilon_0 \over 2T}\right. \nonumber \\
&& \left. + \ i \epsilon_0 \tanh {\epsilon_0 \over 2T} \right],
\label{Y}
\end{eqnarray}
where $R_N=2/e^2v_FN_FS$ is the normal (Sharvin) resistance of the
point contact. Equation (\ref{Y}) agrees with the result obtained
by Averin and Bardas \cite{AB2} in adiabatic regime \cite{eta}.
The quasiparticle conductance is given by the real part of
(\ref{Y}):
\begin{eqnarray}
{G(\omega) \over G_N} &=& {\pi \eta \Omega_0^2 \over (\omega^2 +
4\eta^2)T} \ {\rm sech}^2 {\epsilon_0 \over 2T}, \label{G}
\end{eqnarray}
where $G_N = 1/R_N$. Notice that the right hand side of
Eq.~(\ref{G}) vanishes at $\phi=0$. This however is the point at
which the linear expansion (\ref{gammaexpnd}) fails. One
therefore expects the terms neglected in (\ref{G}) to dominate
$G(\phi=0)$. Similarly, at small $T$, Eq.~(\ref{G}) is
exponentially small (except at $\phi=\pi$). Since in such a
regime, $\omega/T$ becomes large, the expansion of the hyperbolic
tangents in powers of $\omega/T$ will be invalid. Therefore,
deviation from (\ref{G}) at low $T$ is expected. In the next
subsection, we will see such deviations in the numerical results.

The imaginary part of the admittance is also important because it
provides information about the inductive or capacitive behavior of
the junction. Notice that at $\phi=0$, the first term in (\ref{Y})
vanishes but the second term survives. Thus unlike the
conductance, Im$(Y)$ does not vanish; it rather stays finite and
behaves purely inductively ($\sim 1/\omega L$). Similarly, near
$T=0$, the second term in (\ref{Y}) dominates, resulting in a
finite (again inductive) Im$(Y)$. One therefore expects that the
leading order expansion provides good approximation. The
exception is at $\phi=\pi$ where $\epsilon_0=0$ and thus the
second term in (\ref{Y}) vanishes. The higher order terms
therefore play important role in such a case. In the next
subsection, we observe this behavior by comparing with the
numerical results.

\subsection{Numerical Results}

We now present the results of our numerical calculation. The value
of $|\Delta_0|$ is calculated directly from the BCS gap equation
\cite{tinkham}:

\begin{equation}
1 = \lambda \int_{|\Delta_0|}^{\epsilon_c} {d \epsilon \over
\sqrt{\epsilon^2 - |\Delta_0|^2}} \tanh \left( {\epsilon \over
2T} \right),
\end{equation}
where the dimensionless coupling constant $\lambda$ is chosen in
such a way to give $\Delta_0 {\rightarrow} 0$ as $T {\rightarrow}
T_c$. Near $T = 0$, one finds $|\Delta_0| \approx 1.75 T_c$. All
the energy scales ($T,\epsilon,\omega,eV_0,\eta$, etc.) are
normalized to $T_c$ and $Y(\omega)$ to $G_N$. In all calculations,
we take $\eta = 0.01$ and $\epsilon_c = 20$.

\begin{figure}[h]
\epsfysize 4.7cm \epsfbox[50 230 400 510]{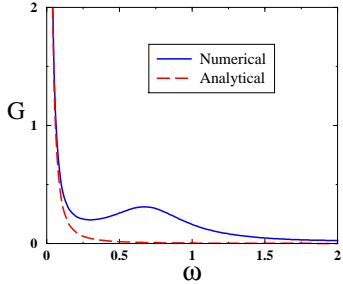} \caption{
Linear conductance $G$ (normalized to $G_N$) as a function of
$\omega$, for $\phi=3\pi/4$ at $T=0.1$. ($\omega$ and $T$ as well
as other energy scales are normalized to $T_c$.)} \label{Gomg}
\end{figure}

\begin{figure}[h]
\epsfysize 7.5cm \epsfbox[80 100 400 500]{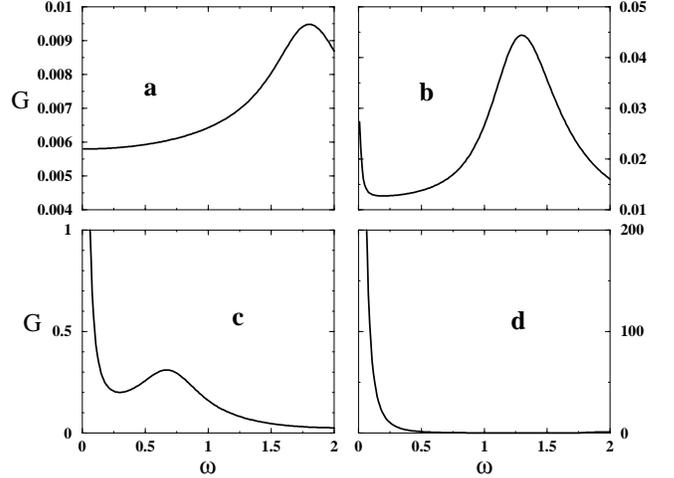}
\caption{Linear conductance as a function of $\omega$ at $T=0.1$.
The phase differences are (a) $\phi=0$, (b) $\phi=\pi/2$, (c)
$\phi=3\pi/4$, and (d) $\phi=\pi$.} \label{RIomg}
\end{figure}

Fig.~\ref{Gomg} compares the result of numerical calculation of
$G$ at $\phi=3\pi/4$ with the analytical result obtained from
Eq.~(\ref{G}). As expected, the two curves overlap at low
frequencies but deviate at larger $\omega$. Around $\omega=0$,
there exists a sharp peak corresponding to the Lorentzian
$\omega$-dependence in (\ref{G}). At larger frequencies, a second
peak appears in the numerical curve which is absent in the
analytical one. The peak clearly results form the higher order
contributions which were neglected in derivation of (\ref{Y}).
Fig.~\ref{RIomg} displays numerical $G$-$\omega$ curves at
different phase differences. At $\phi=\pi$ (Fig.~\ref{RIomg}d),
the sharp peak at $\omega=0$ has the largest value. At smaller
phase differences, this peak becomes less pronounced and
eventually disappears at $\phi=0$, as it should according to
(\ref{G}); the equation actually predicts zero conductance,
therefore the small conductance in Fig.~\ref{RIomg}a is
completely due to the terms neglected in (\ref{G}). The second
peak, however, appears at $\omega= |\Delta_0| \approx 1.75$ for
$\phi=0$ (Fig.~\ref{RIomg}a), and moves towards $\omega=0$ as
$\phi \rightarrow \pi$. It is easy to see that the peak is always
at $\omega = \epsilon_0(\phi)$, i.e., the energy of the Andreev
levels.

\begin{figure}[h]
\epsfysize 4.5cm \epsfbox[50 270 400 510]{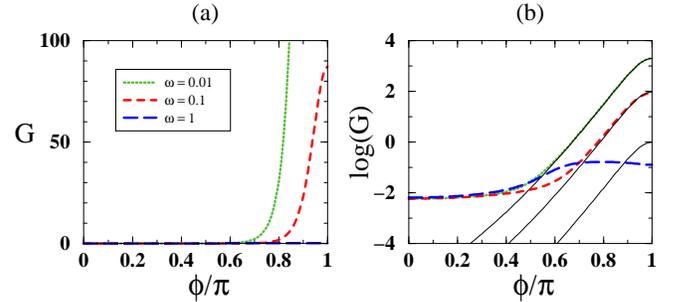} \caption{(a)
Linear conductance $G$ as a function of the phase difference at
$T=0.1$, for different frequencies $\omega$. (b) The same data on
logarithmic scale. The solid lines correspond to the analytical
results. Here (and in the figures that follow), the legend is
common between (a) and (b). } \label{RIph}
\end{figure}

Fig.~\ref{RIph} displays the conductance $G$ as a function of the
phase difference across the contact for different frequencies. It
is clear from the figure that $G$ is strongly phase-dependent.
Especially, it is sharply peaked close to $\phi=\pi$ [for
$\omega=0.01$, it is more than five orders of magnitude larger
than $G(\phi{=}0)$, see Fig.~\ref{RIph}b]. The strong conductance
can be attributed to the existence of zero energy Andreev bound
states (ZBS) [i.e. $\epsilon_0(\phi{=}\pi)=0$]; they provide
large density of states at zero energy. A comparison with the
analytical results is shown, in logarithmic scale, in
Fig.~\ref{RIph}b. As expected, the agreement between the two
calculations at $\omega=0.01$, is good near $\phi=\pi$ but they
deviate as $\phi \rightarrow 0$. The curves however overlap less
at higher frequencies. Especially, at $\omega=1$ they show
completely different phase dependence.

\begin{figure}[h]
\epsfysize 4.5cm \epsfbox[70 270 400 510]{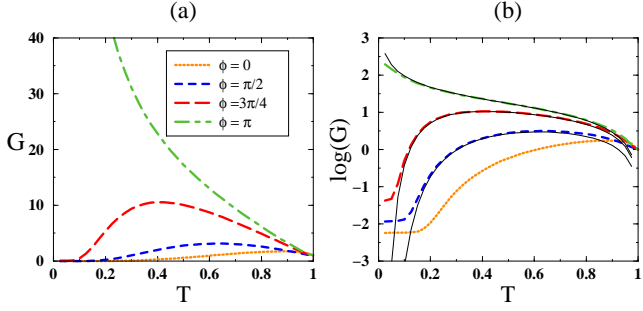} \caption{(a)
Linear conductance as a function of temperature for different
phase differences at $\omega=0.1$. (b) The same data plotted in
logarithmic scale. Solid lines are analytical results. Notice
that at $\phi=0$, Eq.~(\ref{G}) gives $G=0$, therefore no
analytical curve is shown in the figure.} \label{RIT}
\end{figure}

The temperature dependence of the linear conductance is presented
in Fig.~\ref{RIT}a. All the curves join at $G=1$ (or $G=G_N$
before normalization) as $T {\rightarrow} T_c$. This indeed is
expected, because at $T=T_c$ the superconductor becomes normal. It
is clear from the figure that the conductance behaves completely
differently at $\phi = \pi$ compared to other phase differences.
At $\phi=\pi$, the conductance grows with lowering the
temperature and exhibits a $1/T$ dependence in agreement with
(\ref{G}). This type of $1/T$ behavior also exists in the dc
Josephson current at $\phi=\pi$, and is associated with the
current carried by the ZBS. For $\phi=0,\pi/2$, and $3\pi/4$, and
at intermediate temperatures, the linear conductance behaves as
$G \sim e^{-\epsilon_0(\phi)/T}$, in agreement with
Eq.~(\ref{G}). This form of suppression resembles the thermal
activation behavior ($G \sim e^{-\Delta/T}$) in tunnel junctions
\cite{tinkham}. At lower temperatures however, a deviation from
such a behavior occurs. To examine this more carefully, and also
to compare with the analytical results, we have plotted the same
graph in logarithmic scale in Fig.~\ref{RIT}b, adding to it the
analytical curves (solid lines). Except for the $\phi=0$ case
[where (\ref{G}) vanishes], the agreement between the numerical
and analytical results at intermediate temperatures is very good.
At low $T$, on the other hand, the numerical curves show
saturation. The crossover temperature to the saturation regime is
proportional to $\epsilon_0(\phi)$ and is almost independent of
$\omega$ (not shown in the figure).

\begin{figure}[t]
\epsfysize 4.5cm \epsfbox[70 270 400 510]{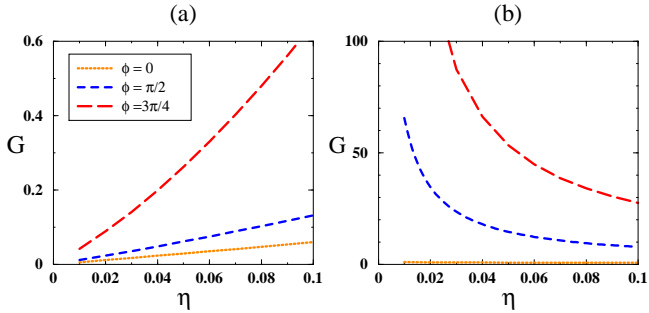} \caption{(a)
Conductance $G$ as a function of damping rate $\eta$ at
$\omega=0.005$. (a) $T=0.01$, \ (b) $T=0.5$. } \label{G-eta}
\end{figure}

Such a saturation does not occur in the analytical curves
(naturally, was also not predicted in Ref.~\cite{AB2}), and is
clearly a higher order property. As we mentioned before, the
leading order result of (\ref{G}) vanishes as $T \rightarrow 0$,
therefore the only remaining contribution will be the higher
order terms. To see this explicitly, in Fig.~\ref{G-eta} we have
plotted $G$ versus $\eta$ at a low frequency ($\omega=0.005$) and
for two different temperatures. One immediately notices
significant difference in the $\eta$-dependence between the two
cases. At high $T$ (Fig.~\ref{G-eta}b), the conductance follows
$1/\eta$ dependence in agreement with (\ref{G}). At $T=0.01$
(Fig.~\ref{G-eta}a), on the other hand, all three curves show
linear dependence on $\eta$, which is obviously higher order than
$1/\eta$. Physically, the residual conductance is a result of the
overlap of the midgap states, broadened by finite $\eta$, at zero
energy. Increasing $\eta$, increases the density of states at
zero energy and therefore the conductance. In reality, $\eta$ is
also temperature dependent and in $s$-wave superconductors, it
vanishes at $T=0$, and so does $G$.

\begin{figure}[h]
\epsfysize 4.5cm \epsfbox[70 260 400 510]{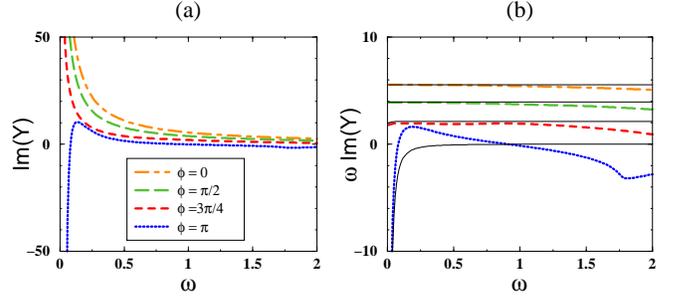} \caption{(a)
The imaginary part of the admittance as a function of $\omega$ at
$T=0.1$. (b) The same data multiplied by $\omega$ together with
the analytical (solid) curves.}\label{IIomg}
\end{figure}

Fig.~\ref{IIomg}a shows the frequency dependence of Im($Y$) for
different phase differences. Except for the $\phi = \pi$ curve,
the other curves seem to show a $1/\omega$ form (inductive
behavior). To see this more clearly, we have plotted
$\omega$Im($Y$) as a function of $\omega$ in Fig.~\ref{IIomg}b.
The first three curves are almost constant, confirming the
$1/\omega$ dependence. The curve at $\phi=\pi$, on the other
hand, exhibits completely different behavior. In
Fig.~\ref{IIomg}b, we also present the analytical curves
corresponding to Eq.~(\ref{Y}). The agreement between the
numerical and analytical curves is good at low frequencies. At
higher frequencies, all curves deviate from their analytical
counterparts, as expected. Especially, for the case of $\phi=\pi$
the discrepancy between the two curves is significant.

\begin{figure}[h]
\epsfysize 4.7cm \epsfbox[60 230 400 490]{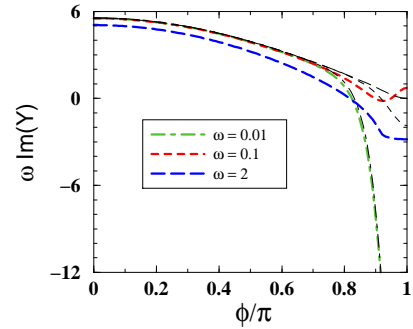}
\caption{$\omega$ Im($Y$) as a function of phase difference for
different frequencies at $T=0.1$. The thin lines correspond to the
low-frequency analytical results.} \label{IIph}
\end{figure}

To understand this better, we have plotted $\omega$Im($Y$) versus
the phase difference $\phi$ for different frequencies in
Fig.~\ref{IIph}. The thin curves are plotted using Eq.~(\ref{Y}).
For $\omega=0.01$ and 0.1, the curves overlap and agree quite well
with the analytical ones over a wide range of $\phi$. Near
$\phi=\pi$, the curves separate and deviation from the analytical
results become more evident. At high frequency ($\omega=2$) on
the other hand, the deviation already exists at $\phi=0$ and
increases as $\phi \rightarrow \pi$.

\begin{figure}[h]
\epsfysize 5cm \epsfbox[60 240 400 510]{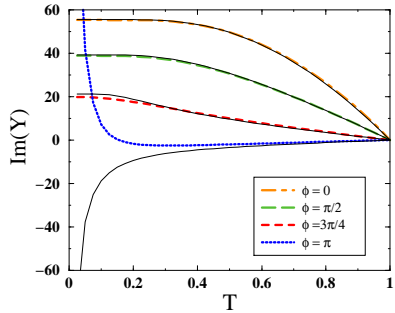} \caption{Im($Y$)
as a function of temperature for different phase differences at
$\omega=0.1$. The solid curves show the corresponding analytical
results.} \label{IIT}
\end{figure}

The temperature dependence of Im($Y$), for different values of the
phase difference, is plotted in Fig.~\ref{IIT}. All the curves
meet at Im($Y$)=0 as $T \rightarrow T_c$. The agreement with
analytical results is very good at high $T$. At lower
temperatures, the $\phi=\pi$ curve behaves completely differently
than the other curves and deviates significantly from the
analytical curve. This, as we mentioned before, is a result of
the breakdown of the small frequency expansion at low
temperatures.

\section{Conclusions}

We have presented a microscopic formalism for calculating ac
properties of Josephson junctions. The method is based on linear
response treatment of the non-equilibrium quasiclassical Green's
function theory of superconductivity and uses a generalized form
of the Riccati-transformations. The self-consistency equation for
the linear response part of the order parameter, as well as the
charge neutrality condition within the bulk, is satisfied to a
very good approximation, with no need for numerical iteration.

We successfully applied the method to the case of a ballistic
superconducting point contact and obtained nontrivial results for
linear conductivity of the junction both analytically and
numerically. We noticed strong temperature, frequency, and phase
dependence in the real and imaginary parts of the ac
conductivity. In particular, we found the conductance to be many
orders of magnitude larger at $\phi=\pi$ than at smaller phase
differences. This is a result of the influence of the zero energy
bound states on the quasiparticle conductance. The agreement
between the analytical and numerical results is very good at low
frequencies. The exceptions happen near $\phi=0$ for $G$, and
$\phi=\pi$ for Im$(Y)$, where the leading order contributions
vanish or become comparable to the neglected terms. The
discrepancy becomes more pronounced at low $T$ or high $\omega$,
where the validity of the leading order approximation become
questionable.

Experimentally, superconducting point contacts have been realized
using techniques such as: scanning tunneling microscopy
\cite{STM}, mechanically controllable break junctions
\cite{Muller,Vleeming}, superconductor--2 dimensional electron
gas--superconductor junctions \cite{Takayanagi,Mur}, etc. Subgap
structures were observed \cite{sgap} and nice measurements of
transmission coefficients of individual quantum channels were
performed by Scheer {\em et al.} \cite{Scheer}. Unfortunately,
phase-dependent measurement of the conductance is difficult and
the only available data (to our knowledge) is those by Rifkin and
Deaver \cite{RD}. They found a strongly phase dependent
conductance, in qualitative agreement with our results. More
experimental research is necessary to confirm the predictions of
the present work.

In this article, we have only considered a contact with perfect
transparency. The method, however, is general and applicable for
arbitrary transparency. One only needs to take into account
appropriate (e.g. Zaitsev) boundary conditions at the contact. A
solution to the Zaitsev boundary conditions, suitable for the
present calculation method, is given in Ref.~\cite{eschrig2}. The
method proposed in this article is also applicable to other
systems such as grain boundary junctions between unconventional
superconductors, which is the subject of a separate publication
\cite{dwave}.

\section*{Acknowledgement}

The author has greatly benefited from discussion with A.~Leggett,
A.~Omelyanchouk, A.~Smirnov, A.~Zagoskin, and especially
A.~Maassen van den Brink because of helpful suggestions in the
preparation of the manuscript and providing Eq.~(A14) as the
appropriate form for the gauge transformation.

\appendix
\section{Quasiclassical Keldysh Green's functions}

The quasiclassical Green's function \cite{rammer} $\breve{g}({\bf
v}_{F},\epsilon ;{\bf r},t)$ is a 2$\times$2 matrix, every element
of which also a 2$\times$2 matrix, and a function of the Fermi
velocity ${\bf v}_{F}$, quasiparticle energy $\epsilon$, position
${\bf r}$, and time $t$. We represent $\breve{g}$ as

\begin{equation}
\breve{g}= \left(
\begin{array}{cc}
\widehat{g}^R & \widehat{g}^K \\
0 & \widehat{g}^A
\end{array}
\right), \label{g}
\end{equation}
where the matrices $\widehat{g}^R,\ \widehat{g}^A,\ \widehat{g}^K$
are the quasiclassical retarded, advanced, and Keldysh Green's
functions in Nambu-Gorkov representation, respectively:

\begin{equation}
\widehat{g}^{R,A}=\left(
\begin{array}{cc}
g^{R,A} & f^{R,A} \\
f^{R,A\dagger} & -g^{R,A\dagger}
\end{array}
\right) , \quad \widehat{g}^K=\left(
\begin{array}{cc}
g^K & f^K \\
-f^{K\dagger} & g^{K\dagger}
\end{array}
\right). \label{defg}
\end{equation}
The $\dagger$-operation performs the following transformation

\begin{equation}
{\cal O}^\dagger({\bf v}_F,\epsilon;{\bf r},t) = {\cal O}(-{\bf
v}_F,-\epsilon;{\bf r},t)^*. \label{odagger}
\end{equation}
In frequency domain it also changes $\omega$ to $-\omega$. The
retarded and advanced Green's functions carry information about
the energy spectrum of the electronic states of the system, while
the Keldysh ones have information about occupation of those
states.

The following symmetries hold for the Green's functions:
\begin{equation}
g^A = - g^{R*} ,\qquad f^{R\dagger} = f^{A*}, \label{symRA}
\end{equation}
and

\begin{equation}
g^K  = g^{K*} ,\qquad f^{K\dagger} = f^{K*}. \label{symK}
\end{equation}
In frequency space, these give $g^A(\omega) = - g^{R}(-\omega)^*$,
etc.

The equation of motion that describes the time evolution of
$\breve{g}$ is written as

\begin{eqnarray}
{\bf v}_{F}\cdot \nabla \breve{g} - \left[i \left( \epsilon -
{e\over c} {\bf v}_F \cdot {\bf A} \right) \breve {\tau }_{3} -
\breve{\Delta} + ie \Phi \breve{1},\ \breve{g} \right]_\otimes =
\breve{0} \label{QCEq}
\end{eqnarray}
accompanied by the normalization condition

\begin{eqnarray}
\breve{g} \otimes \breve{g} = \breve{1}, \label{norm}
\end{eqnarray}
where $\Phi$ and ${\bf A}$ are the scalar and vector potentials
respectively, and \ ${[A,B]}_\otimes \equiv A \otimes B - B
\otimes A$, with

\begin{equation}
(A \otimes B )(\epsilon,t) = e^{{i \over 2}
\left(\partial^A_\epsilon \partial^B_t - \partial^A_t
\partial^B_\epsilon \right)} A(\epsilon,t) B(\epsilon,t).
\label{otimes}
\end{equation}
The product (\ref{otimes}) is associative and satisfies $(A
\otimes B)^\dagger = A^\dagger \otimes B^\dagger$, but $(A
\otimes B)^* = B^* \otimes A^*$. We also have

\begin{equation}
\breve{\tau}_3 = \left(
\begin{array}{cc}
\widehat{\tau}_3 & 0   \\
0 & \widehat{\tau}_3
\end{array}
\right), \qquad
\breve{\Delta}= \left(
\begin{array}{cc}
\widehat{\Delta} & 0  \\
0 & \widehat{\Delta}
\end{array}
\right),
\end{equation}
where the 2$\times$2 Pauli matrix $\widehat{\tau}_3$ and
$\widehat{\Delta}$ are

\begin{equation}
\widehat{\tau}_3= \left(
\begin{array}{cc}
1 & 0  \\
0 & -1
\end{array}
\right), \qquad
\widehat{\Delta}= \left(
\begin{array}{cc}
0 & \Delta  \\
\Delta^\dagger & 0
\end{array}
\right),
\end{equation}
with $\Delta$ being the superconducting pairing potential. The
constant $e$ in (\ref{QCEq}) is the absolute value of the
electronic charge, and $c$ is the speed of light. In equilibrium,
Eq.~(\ref{QCEq}) reduces to the Eilenberger equation (see
Ref.~\cite{amin} for example), by changing $\epsilon \rightarrow
i\omega_n$, where $\omega_n$ are the Matsubara frequencies.

A gauge transformation is defined by the following simultaneous
operations:

\begin{eqnarray}
\breve{\Delta} &\mapsto& e^{i (\chi/2) \breve{\tau}_3}
\breve{\Delta} e^{-i (\chi/2) \breve{\tau}_3}, \\
\Phi &\mapsto& \Phi + {1 \over 2e} \partial_t \chi, \\
{\bf A} &\mapsto& {\bf A} - {c \over 2e} \nabla \chi, \\
\breve{g} &\mapsto& e^{i (\chi/2) \breve{\tau}_3} \otimes
\breve{g} \otimes e^{-i (\chi/2) \breve{\tau}_3}.
\end{eqnarray}
It therefore takes the phase of the order parameter $\phi$ to
$\phi + \chi$. In the discussion of section II, we have chosen
$\chi = - \delta \phi$.

The Green's functions can be used to calculate physical
quantities. The quasiparticle density of states is given by

\begin{equation}
N(\epsilon) = {N_F \over 4} \left< {\rm Tr} [\widehat{\tau}_3
(\widehat{g}^R - \widehat{g}^A)] \right>,
\end{equation}
where $N_F$ is the density of state of electrons at the Fermi
surface and $\left< ... \right>$ denotes averaging over ${\bf
v}_F$. The pairing potential satisfies the following
self-consistency relation:

\begin{equation}
\widehat{\Delta}({\bf v}_F) = {N_F \over 4i}
\int_{-\epsilon_c}^{\epsilon_c} d\epsilon \left< V({\bf v}_F,{\bf
v}_F') \widehat{f}^K ({\bf v}_F') \right>_{{\bf v}_F'},
\label{sce}
\end{equation}
where $\widehat{f}^K$ is the off-diagonal part of $\widehat{g}^K$,
and $V({\bf v}_F,{\bf v}_F')$ is the interaction potential.
Furthermore, the current density is given by

\begin{equation}
{\bf j} = {e N_F \over 4} \int_{-\epsilon_c}^{\epsilon_c}
d\epsilon \left< {\bf v}_F\ {\rm Tr}[ \widehat{\tau}_3
\widehat{g}^K ] \right>, \label{j}
\end{equation}
and the charge density by

\begin{equation}
\rho = eN_F \left( -2e\Phi + {1 \over 4}
\int_{-\epsilon_c}^{\epsilon_c} d\epsilon \left< {\rm Tr}
[\widehat{g}^K ] \right> \right). \label{rho}
\end{equation}

\section{Generalized Riccati-transformation}

Riccati-transformations \cite{Schopohl} are proven to be very
useful tools for numerical calculations of equilibrium properties
of superconducting systems. For non-equilibrium systems, however,
the presence of the $\otimes$-operators makes the formalism
nontrivial. Nevertheless, a generalization of the standard
transformation to the non-equilibrium case is possible. It is
common to define Riccati amplitudes $a^\alpha$ and $b^\alpha$,
where $\alpha{=}R,A$ for the retarded and advanced Green's
functions respectively. They are related to the corresponding
Green's functions by \cite{eschrig,eschrig2}

\begin{eqnarray}
g^\alpha &=& s^\alpha (1 + a^\alpha \otimes b^\alpha)^{-1} \otimes
(1 - a^\alpha \otimes b^\alpha), \nonumber \\
f^\alpha &=& s^\alpha (1 + a^\alpha \otimes b^\alpha)^{-1}
\otimes (2 a^\alpha), \label{gfRA}
\end{eqnarray}
where $s^\alpha$ is defined in Eq.~(\ref{s}). Here we define the
inverse operation by $A^{-1} \otimes A = A \otimes A^{-1} = 1$.
Eqs.~(\ref{gfRA}) already resemble their counterparts in the
standard Matsubara formalism \cite{Schopohl}. It is
straightforward to show that for ${\bf A}=0$ these functions
satisfy Riccati-type equations given by

\begin{eqnarray}
{\bf v}_{F}\cdot \nabla a^\alpha &=& 2i \epsilon a^\alpha -
a^\alpha \otimes \Delta^\dagger \otimes a^\alpha +  \Delta +
\left[ ie\Phi , a^\alpha \right]_\otimes,  \nonumber \\
-{\bf v}_{F}\cdot \nabla b^\alpha &=& 2i \epsilon b^\alpha -
b^\alpha \otimes \Delta \otimes b^\alpha + \Delta^\dagger -
\left[ ie\Phi , b^\alpha \right]_\otimes \label{dab}
\end{eqnarray}
In equilibrium, the $\otimes$-operation is replaced by a simple
multiplication and these equations reduce to (\ref{rcti}).

It is also necessary to define other functions $a^K$ and $b^K$
\cite{abk}, which are related only to the Keldysh Green's
functions \cite{eschrig,eschrig2}

\begin{eqnarray}
g^K &=& 2 (1 + a^R \otimes b^R)^{-1} \otimes (a^K + a^R \otimes
b^K \otimes b^A) \nonumber \\
&& \otimes \ (1 + a^A \otimes b^A)^{-1}, \nonumber \\
f^K &=& 2 (1 + a^R \otimes b^R)^{-1} \otimes (a^K \otimes a^A -
a^R \otimes b^K) \nonumber \\
&& \otimes \ (1 + b^A \otimes a^A)^{-1}, \label{gfK}
\end{eqnarray}
and are governed by the following dynamical equations

\begin{eqnarray}
{\bf v}_{F}\cdot \nabla a^K &=& - \partial_t a^K - a^R \otimes
\Delta^\dagger \otimes a^K + a^K \otimes \Delta \otimes b^A
\nonumber \\ && + \left[ ie\Phi , a^K \right]_\otimes, \nonumber \\
-{\bf v}_{F}\cdot \nabla b^K &=& - \partial_t b^K - b^R \otimes
\Delta \otimes b^K + b^K \otimes \Delta^\dagger \otimes a^A
\nonumber \\ && - \left[ ie\Phi , b^K \right]_\otimes.
\label{dabK}
\end{eqnarray}
The functions $a^\alpha$ and $b^\alpha$ are related, by the
$\dagger$-operation [Eq.~(\ref{odagger})], through
$b^\alpha{=}a^{\alpha \dagger}$ for $\alpha = R,A,K$. In
addition, the symmetries (\ref{symRA}) and (\ref{symK}) require

\begin{equation}
a^A= - b^{R*}, \quad a^K = a^{K*}.
\end{equation}
One can show that Eqs.~(\ref{gfRA})--(\ref{dabK}) satisfy the
dynamical equation (\ref{QCEq}) together with the normalization
condition (\ref{norm}).

In equilibrium, Eqs.~(\ref{gfRA}) reduce to (\ref{gf0RA}), and
(\ref{gfK}) give

\begin{eqnarray}
g_0^K &=& 2 {a_0^K + a_0^R b_0^K b_0^A \over (1 + a_0^R b_0^R)(1 + a_0^A b_0^A)}, \\
f_0^K &=& 2 {a_0^K a_0^A - a_0^R b_0^K \over (1 + a_0^R b_0^R)(1 +
a_0^A b_0^A)}.
\end{eqnarray}
Satisfying (\ref{gKRA}), one finds

\begin{eqnarray}
a_0^K = (1 + a_0^R b_0^A) {\cal F} , \qquad b_0^K = -(1 + a_0^A
b_0^R) {\cal F}. \label{abKRA}
\end{eqnarray}

Linear response equations (\ref{dgRA}) and (\ref{dfRA}) are
obtained by expanding (\ref{gfRA}) to the linear order. To obtain
(\ref{dgX}) and (\ref{dfX}), we expand (\ref{gfK}) to the linear
order and introduce

\begin{equation}
\delta \widehat{g}^X=\left(
\begin{array}{cc}
\delta g^X & \delta f^X \\
\delta f^{X \dagger} & -\delta g^{X \dagger}
\end{array}
\right) \label{mdgx}
\end{equation}
through (\ref{gx}). We also define $\delta a^X$ and $\delta b^X$
in terms of the linear response $\delta a^K$ and $\delta b^K$ by

\begin{eqnarray}
\delta a^K &=& \delta a^X ({\cal F}_+ - {\cal F}_-) + \delta a^R
b_{0-}^A {\cal F}_- + \delta b^A a_{0+}^R {\cal F}_+ , \nonumber \\
\delta b^K &=& \delta b^X ({\cal F}_- - {\cal F}_+) - \delta b^R
a_{0-}^A {\cal F}_- - \delta a^A b_{0+}^R {\cal F}_+ . \label{abx}
\end{eqnarray}
The differential equations (\ref{DifEqs}) then follow directly
from (\ref{dab}) and (\ref{dabK}). It should finally be mentioned
that the $A$'s and $B$'s in Eq.~(\ref{DifEqs}) are related by:
$\widetilde{A}^\alpha = A^{\alpha \dagger}$,
$\widetilde{B}^\alpha = B^{\alpha \dagger}$.


\begin{thebibliography}{99}
\bibitem{dlt} In general, a large voltage applied to a Josephson
junction may affect the modulus of the order parameter. This
however is not a concern in the linear response limit.
\bibitem{jac} In the superconducting point contact considered
here, $j_{\rm c}(r)/j_{\rm c, bulk} \sim (d/r)^2$, where $r$
($\sim \xi_0 \gg d$) is the distance from the orifice and $d$ is
the diameter of the hole (see Fig.~\ref{fig1}).
\bibitem{eta0} Here we include quasiparticle damping
phenomenologically by introducing $\eta$. A more rigorous
calculation should take into account electron-phonon interaction
directly, as the main cause for inelastic scattering.
\bibitem{eta} In order to obtain Eq.~(7) of Ref.~\cite{AB2}, one should
replace $2\eta$ with $\gamma (\epsilon_0)$.
\bibitem{abk} These are indeed equivalent to the distribution
functions $x^K$ and $\widetilde{x}^K$ defined in
Refs.~\cite{eschrig,eschrig2}.
\end{thebibliography}
\end{document}